\documentclass[lettersize,journal]{IEEEtran}
\usepackage{amsmath,amsfonts}
\usepackage{algorithmic}
\usepackage{algorithm}
\usepackage{array}
\usepackage[caption=false,font=normalsize,labelfont=sf,textfont=sf]{subfig}
\usepackage{textcomp}
\usepackage{stfloats}
\usepackage{url}
\usepackage{verbatim}
\usepackage{graphicx}
\usepackage{cite}
\usepackage{color}
\usepackage{multirow}
\usepackage[colorlinks,linkcolor=black,anchorcolor=black,citecolor=black,CJKbookmarks=True]{hyperref}
\usepackage[numbers]{natbib}
\usepackage{booktabs}

\begin{document}

\title{{Knowledge-aware Diffusion-Enhanced Multimedia Recommendation}}

\author{Xian Mo, Fei Liu, Rui Tang, Jintao, Gao, Hao Liu, \textit{Member, IEEE}
        \thanks{Manuscript received Xx xx, 2024; revised xx xx, 2024. Our work was funded in part by the Natural Science Foundation of Ningxia (2024AAC05011), the National Natural Science Foundation of China (62306157, 62202320, 62462051), and the Natural Science Foundation of Sichuan Province (2024NSFSC1449).}

        \IEEEcompsocitemizethanks{\IEEEcompsocthanksitem Xian Mo, Fei Liu, Jintao Gao, and Hao Liu were with the School of Information Engineering, Ningxia University, Yinchuan, 750021, China (E-mail: mxian168@nxu.edu.cn; liufei0206@stu.nxu.edu.cn; gaojintao@nxu.edu.cn; liuhao@nxu.edu.cn); Rui Tang was with School of Cyber Science and Engineering, Sichuan University, Chengdu 610065, Sichuan, China (E-mail: tangrscu@scu.edu.cn).

}

}



\maketitle

\begin{abstract}
Multimedia recommendations aim to use rich multimedia content to enhance historical user-item interaction information, which can not only indicate the content relatedness among items but also reveal finer-grained preferences of users.
In this paper, we propose a \underline{K}nowledge-aware \underline{Diff}usion-\underline{E}nhanced architecture using contrastive learning paradigms (KDiffE) for multimedia recommendations.
Specifically, we first utilize original user-item graphs to build an attention-aware matrix into graph neural networks, which can learn the importance between users and items for main view construction. The attention-aware matrix is constructed by adopting a random walk with a restart strategy, which can preserve the importance between users and items to generate aggregation of attention-aware node features.
Then, we propose a guided diffusion model to generate strongly task-relevant knowledge graphs with less noise for constructing a knowledge-aware contrastive view, which utilizes user embeddings with an edge connected to an item to guide the generation of strongly task-relevant knowledge graphs for enhancing the item's semantic information.
We perform comprehensive experiments on three multimedia datasets that reveal the effectiveness of our KDiffE and its components on various state-of-the-art methods.
Our source codes are available\footnote{\url{https://github.com/1453216158/KDiffE}}.

\end{abstract}

\begin{IEEEkeywords}
Multimedia recommendation, Knowledge-aware enhanced, diffusion model, contrast learning
\end{IEEEkeywords}

\section{Introduction}
\label{introduction}
Multimedia recommendations (MMRec) \cite{LiuXLSH24} try to additionally mine multimodal user preference cues from multimedia content (e.g., visual, textual, and acoustic) of items as supplement content to enhance item's semantic information, which can indicate content relatedness among items, reveal users’ finer-grained preferences, and improve recommendation performance.
It has been widely used in different real-world applications, such as recipe-related applications \cite{MinJJ20}, content-sharing platforms \cite{Yu0LB23}, and E-commerce \cite{LiuXLSH24}.

In general, the MMRec paradigm consists of broadly two steps.
Specifically, multimodal features are first extracted from multimedia content using pre-trained deep networks \cite{LiuZNSHLFYHM18} and then incorporated into recommendation frameworks to model additional user and item preferences.
Recent research on multimedia recommendation frameworks focused on encoding users and items into low-dimensional vector representations using graph neural networks (GNN) \cite{XieHRSK23} to model user and item preferences, which can learn higher-order relationships between users and items by message propagation mechanisms, thus further improving the representations of users and items.
For example, Cai et al. \cite{CaiQFHX22} present an adaptive multi-modal anti-bottleneck GNN for personalized micro-video recommendation, while MHGCF \cite{LiuXLSH24} introduces three types of GNN to model collaborative signals, content-level preferences, and semantic-level preferences for multimedia recommendation.
Nevertheless, the above existing approaches overlook the different importance between user and item interactions. Approaches based on GNN averagely aggregate all user and item interactions, which may generate inaccurate user and item representations and worsen the multimedia recommendation performance.

{Some multimedia recommendation approaches based on adaptive training attention weight mechanisms \cite{WangWYWSN23, LakmalPBK24} appear, which can learn the different importance between the user and item interactions.}
For example, DualGNN \cite{WangWYWSN23} models the user’s attention on different modalities to learn the multi-modal user preference for micro-video recommendations, while MGCN \cite{Yu0LB23} presents a multi-view GNN to extract modality-shared features via attention mechanisms for multimedia recommendations.
Recently, MONET \cite{KimKSK24} has designed two core components containing both target-aware attention and modality-embracing GNN in multimedia recommender systems.
However, most existing approaches update model parameters by adaptive training attention weight mechanisms, which overlooks constructing an attention-aware module using the topology relationship of user-item graphs into GNN to identify the importance between the user and item interactions for guiding node aggregation, resulting in a high time complexity~\cite{BahdanauCSBB16}.
In more detail, GNN-based approaches mainly employ the Laplacian matrix~\cite{Terroso-SaenzAM23} to generate node aggregation, which overlooks employing user-item graph topology relationship to build an attention-aware matrix into the Laplacian matrix to generate attention-aware node aggregation, resulting in only revealing the connection relationship between users and items, but not the importance between users and items.
Although some approaches learn the importance of user and item interactions by adaptive training attention weight mechanisms~\cite{LakmalPBK24}, this mechanism for downstream network analysis tasks can cause a high computational cost~\cite{BahdanauCSBB16} and is not intuitively understandable~\cite{StaceyBR22}.
From the above discussion, we can infer \textit{how to employ topology relationships of user-item graphs to construct an attention-aware matrix into GNN to learn the importance between user-item interactions to enhance explanation and efficiency remains as a challenge.}

\begin{figure}
  \centering
  \includegraphics[width=0.63 \linewidth]{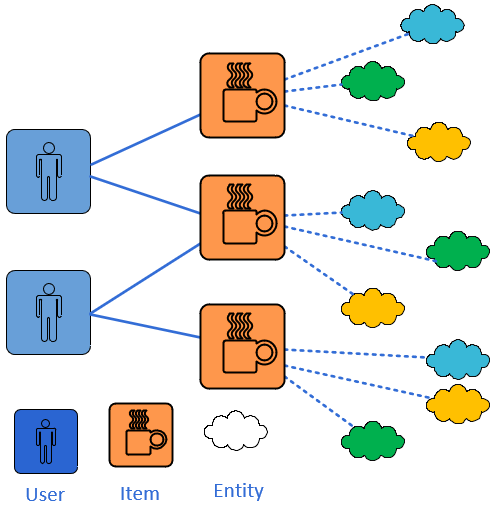}
    \vspace{-5pt}
  \caption{{A simple illustrative example: multimedia recommendation network}}
  \label{toyexample1}
\end{figure}

{To effectively extract multimodal features from multimedia content, some works \cite{LiuXLSH24, JiangYXH24} utilize multimedia content of items as entities and items as nodes to construct a knowledge graph (KG) to enhance items' semantic information.}
A multimedia recommendation network as shown in Figure  \ref{toyexample1} includes users, items, and entities.
Different types of entities represent different types of multimedia content of items, which are distinguished by different colors.
The different types of entities and items can construct a knowledge graph to enhance user-item interaction information, which can reveal users’ finer-grained preferences, indicate content relatedness among items, and improve recommendation performance.
However, multimedia content of items usually suffers from sparsity in real application scenarios, resulting in constructed KG having a limited ability to improve recommendation performance.

To solve the sparsity problem, some researchers have adopted contrastive learning (CL) \cite{ZhangZCLWZ24, JiangYXH24} to generate a new knowledge graph for constructing a knowledge-aware contrastive view.
Specifically, it can generate self-supervised training signals by graph augmentation strategies for alleviating the data sparsity problem.
However, it is unavoidable that multimedia content of items retains plenty of noisy information, which is described by topic-irrelevant connections between entities and items, resulting in irrelevant to user interests.
As a result, the generated contrastive view may be contaminated by multimodal noise, resulting in all user and item representations being explicitly injected with noise information after graph convolution operations, which can degrade recommendation performance.
In recent years, diffusion models (DM) \cite{JiangYXH24, abs-2406-11781} have performed outstanding performance in robust KG generation, which assumes that the original KG follow an unknown probability distribution and tries to approximate that distribution by a neural network to recover the original KG. 
Nevertheless, most approaches ignore utilizing original KG relationships as supplementary content to guide the generation of strongly task-relevant KG with less noise, which can't ensure that the generated KG can always bring benefits towards the strongly task-relevant node knowledge.
From the above discussion, it is obvious that \textit{how to design a guided diffusion model to generate a knowledge-aware contrastive view with less noise for retaining strongly task-relevant node knowledge remains another challenge.}

To address the above two challenges, our work proposes a \underline{K}nowledge-aware \underline{Diff}usion-\underline{E}nhanced architecture using contrastive learning paradigms (KDiffE) for multimedia recommendations as shown in Figure \ref{fig1}.
Specifically, we first use original user-item graphs to construct an attention-aware matrix into graph neural networks to learn the importance between users and items for generating attention-aware node feature aggregation, which is constructed to preserve the importance between users and items by a random walk with a restart strategy~\cite{ZhangSZZ23}.
In more detail, we compute the Jaccard index \cite{WangNB23} for the set of a user and an item sampled by a random walk with a restart strategy as the similarity of the user and item to build the attention-aware matrix.
The attention-aware matrix can clarify the graph embedding propagation layer on the node level with the contributions of each neighboring node.
Thus, the interpretability of our model is enhanced. 
Our model avoids the repetitive updating of parameters in the training process compared with adaptive training attention weight mechanisms.
Hence, it has a lower computational cost.
Then, we propose a guided diffusion model to generate strongly task-relevant knowledge graphs for constructing a knowledge-aware contrastive view, which utilizes user embeddings with an edge connected to an item to guide the generation of task-relevant knowledge graphs for enhancing the item's semantic information.
Furthermore, the generated KG structure is adjusted by top-$q$ relations between items and entities that are strongly task-relevant, which can preserve the informative structure of the reconstructed knowledge graph with less noise.

Our paper makes the following contributions:

\begin{itemize}
  \item We present an effective contrastive learning architecture KDiffE for multimedia recommendations, which design an attention-aware matrix to identify the importance between users and items for generating attention-aware node feature aggregation.
  \item We propose a guided diffusion model to generate a knowledge-aware contrastive view, which utilizes user embeddings with an edge connected to an item to guide the generation of strongly task-relevant node KG with less noise for enhancing node semantic information.

  \item Comprehensive experiments on three multimedia datasets demonstrate the effectiveness of our KDiffE and its components on various state-of-the-art methods.

\end{itemize}

\section{Related Work}

\noindent
\textbf{GNN-based multimedia recommendations.} 
MMRec first adopt pretrained neural networks to extract the multimodal data of items and learn their feature representations.
Then, the learned feature representations are integrated into recommendation frameworks to model additional user preferences.
In recent years, GNN-based MMRec approaches have demonstrated superior performance in learning node representations on graphs, which can learn higher-order affinities by stacking numerous embedded propagation layers.
For example, early works \cite{WangWTSRL17} extract only deep visual features by pre-trained GNN and employ them to enhance item representations.
Later works extract multimodal features and integrate them into item representations using GCN.
MMALFM \cite{ChengCZKK19} extracts images and review features by the proposed multimodal aspect-aware latent factor model to learn user preference, while MMGCN \cite{WeiWN0HC19} extracts different modality features by multiple GNN modules to learn fine-grained user preferences.
Recently, Cai et al. \cite{CaiQFHX22} presented an adaptive multi-modal anti-bottleneck GNN for personalized micro-video recommendation, while MHGCF \cite{LiuXLSH24} extracts collaborative signals, content-level preferences, and semantic-level preferences by constructing three types of GNN, which utilizes multimedia content of items as entities and items as nodes to construct a knowledge graph to enhance items' semantic information.
However, the above existing approaches cannot learn the different importance between user and item interactions.

{Some multimedia recommendation approaches based on adaptive training attention weight mechanisms \cite{TaoWWHHC20, LakmalPBK24} appear to learn the different importance between user and item interactions.}
For example, MGAT \cite{TaoWWHHC20} learns the weight of user preferences over different modalities by constructing additional attention modules, while DualGNN \cite{WangWYWSN23} learns the multi-modal user preference by modeling the user’s attention on different modalities.
Recently, MGCN \cite{Yu0LB23} extracts modality-shared features via attention mechanisms by a multi-view GNN, while MONET \cite{KimKSK24} designs two core components containing both target-aware attention and modality-embracing GNN in multimedia recommender systems.
Adaptive training attention weight mechanisms for downstream network analysis tasks can cause a high computational cost~\cite{BahdanauCSBB16} and are not intuitively understandable~\cite{StaceyBR22}.

\noindent
\textbf{CL-based multimedia recommendations.} 
In real application scenarios, multimedia content of items often suffers from sparsity issues. Some researchers adopt contrastive learning to augment graph data to generate self-supervised training signals on user-item graphs.
For example, Liu et al. \cite{LiuMSO022} learn intra-modal and inter-modal features by a multi-modal contrastive pretraining model, while CLCRec~\cite{WeiWLNLLC21} adopts multi-modal features using contrastive learning to enrich item embeddings for handling the item cold-start problem in MMRec.   
GHMFC \cite{WangWC22} uses graph neural networks to learn multi-modal embeddings for constructing two contrast learning modules, while MMGCL \cite{Yi0OM22} enhances multi-modal representations by modality edge dropout and modality masking in a self-supervised learning manner.
Recently, MICRO \cite{ZhangZLZWW23} learns item-item affinities for each modality by a contrastive modality fusion model, while BCCL \cite{YangFZWL23} adopts a Modal-aware Bias Constrained Contrastive Learning approach to improve the sparse modal feature. 
Later, BM3~\cite{ZhouZLZMWYJ23} adopts self-supervised learning to eliminate the need for randomly sampled negative samples in MMRec, while MGCL~\cite{LiuXGSQH23} learns visual preference clues and textual preference clues using a CL-based strategy in MMRec.

\noindent
\textbf{DM-based recommendations.}
{In recent years, diffusion models \cite{JiangYXH24, abs-2406-11781} have achieved excellent performance in robust graph data generation.
DiffuASR \cite{LiuYZDGT023} reconstructs the embedding sequence matrix by a diffusion-based SU-Net \cite{Ronneberger17} architecture, while  Diff4Rec \cite{Wu0CLHS023} corrupting and reconstructing the user-item interactions to generate diversified augmentations.
More recently, PDRec \cite{MaXMCZLK24} generate the top-ranked unobserved items by a positive augmentation strategy, while DiffMM~\cite{abs-2406-11781} incorporates a cross-modal CL paradigm with a modality-aware graph diffusion model in MMRec.
However, most existing methods focus on traditional non-multimodal recommendations and cannot be directly applied to multimedia recommendations.
Hence, employing diffusion models to generate robust contrastive views for multimedia recommendations is very worthy of study.}

\section{Problem Definition}
In this section, we introduce some essential concepts and give a formal definition of multimedia recommendations.

\smallskip \noindent
\textbf{User-item Graph.} A user-item graph can be defined as $G=(U, V, Y)$, where $U=\{u_1, \dots, u_i,\dots, u_I\}$ with $(|U|=I)$ represents the set of users and $V=\{v_1, \dots, v_j,\dots, v_J\}$ with $(|V|=J)$ represents the set of items, respectively.
$I$ and $J$ represent the number of users and items.
$Y$ defines the interaction matrix between users and items, and $Y=[y_{i,j}]_{I \times J} \in \{0, 1\}$ represent the interaction between user $u_i$ and item $v_j$.
If $y_{ij}=1$, it means there exists an interaction between user $u_i$ and item $v_j$, and $y_{ij}=0$ otherwise.

\smallskip \noindent
\textbf{Knowledge Graph.} A knowledge graph can be represented as $G_k=(h, r, t)$, which is utilized to organize external multimedia content (e.g., visual, textual, and acoustic) by incorporating different types of multimodal features and their corresponding relations.
The semantic relatedness between the head entity $h$ and the tail entity $t$ in relation type $r$ can be defined as triplet $(h, r, t)$, where the head entity $h$ and tail entity $t$ represent items in user-item graphs $G$ and a specific type of multimedia content of items, respectively.
Therefore, we can effectively employ different types of multimedia content as tail entity $t$ as supplementary content to improve the item's semantic information to model additional user preferences.

\smallskip \noindent
\textbf{Multimedia Recommendation.} Given a user-item graph $G=(U, V, Y)$ and the associated knowledge graph $G_k=(h, r, t)$, we first aggregate different types of entities in KG into items $V$ to enhance items' semantic information for constructing a knowledge-aware user-item graph $\textbf{G=(U, V, Y)}$.
We then construct an attention-aware matrix $S$ into GNN to identify the importance between users $\textbf{U}$ and items $\textbf{V}$ for generating attention-aware node feature aggregation, where $S$ can be got by computing the Jaccard index for the set of a user and an item sampled by a random walk with a restart strategy.
Next, we utilize user embeddings with an edge connected to an item to design a guided diffusion model, which can guide the generation of strongly task-relevant knowledge graphs $\hat{G_k}$ with less noise for generating knowledge-aware contrastive view $\mathbf{\hat{G}}$.
Finally, we adopt a contrastive loss function to project each user and item to a low-dimensional vector representation to preserve the topological structure and semantic relations.
Hence, our multimedia recommendation task focuses on predicting the unobserved user-item interaction ${y}_{ij}$ with corresponding user and item representations. 

\begin{figure*}
  \centering
\section{  \includegraphics[width=0.7 \linewidth]{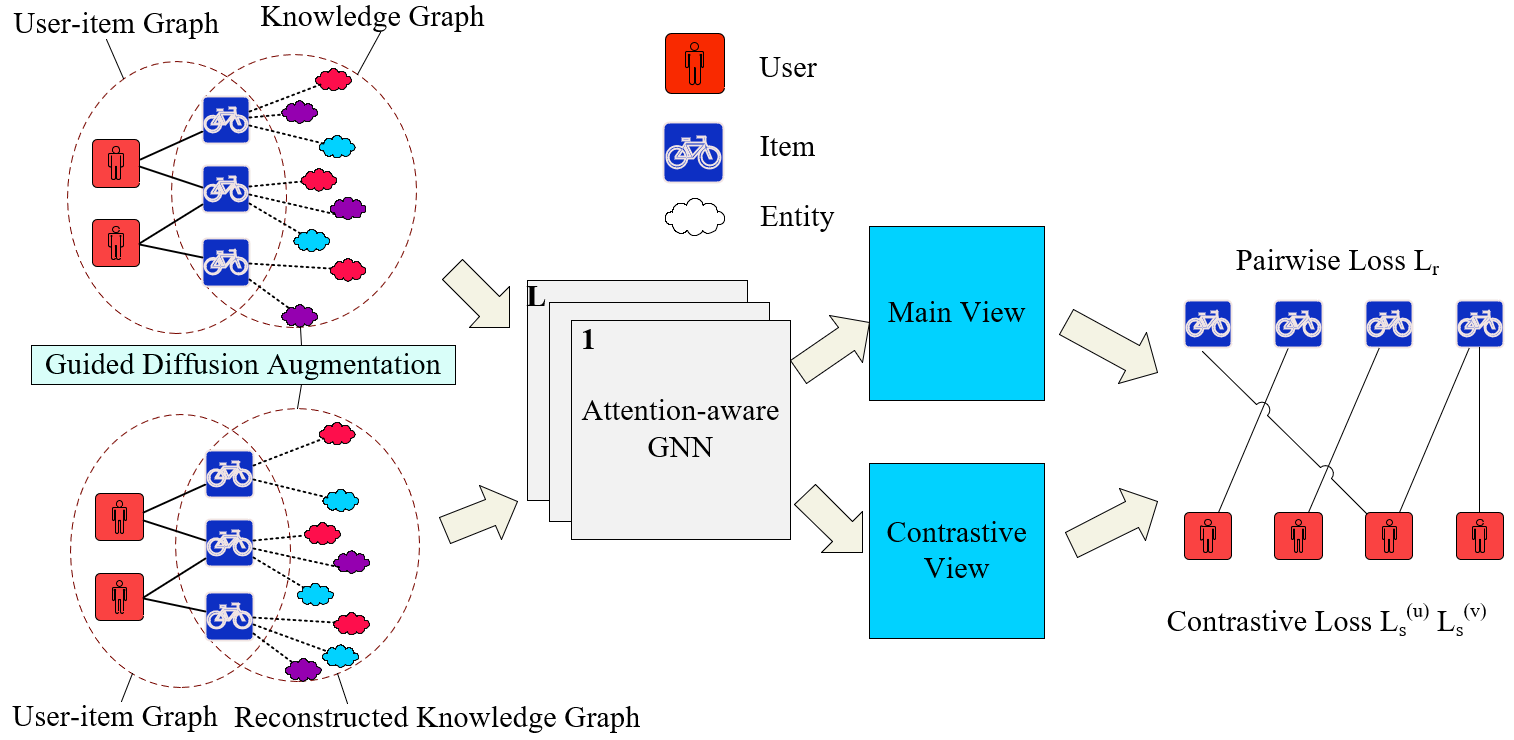}}
    \vspace{-5pt}
  \caption{{Overall architecture of KDiffE model: it first presents a guided diffusion model to generate a KG with less noise for data augmentation; it then presents a graph message passing layers with an attention-aware mechanism for main view and contrastive view construction; a pairwise loss $L_r$ and contrastive loss $L^{(u)}_{s}/L^{(v}_{s}$ (refer to Formulas \ref{formula403} and \ref{contrastive-1} for a precise definition.) will ultimately be employed to contrast augmented view embeddings with main-view embeddings for model parameters training.}}
  \label{fig1}
\end{figure*}

\section{Methodology}
In this section, we introduce the KDiffE model architecture as shown in Figure \ref{fig1}, which consists of two main parts.
The first part presents a graph message passing layers with an attention-aware mechanism to identify the importance between users and items for main view construction.
The second part presents a guided diffusion model to generate a knowledge-aware contrastive view with less noise for data augmentation.

\subsection{User-item Graph Learning with Attention-aware}
In this section, we first employ pre-trained neural networks to extract the multimodal entities for constructing an associated knowledge graph $G_k=(h, r, t)$. 
Then, a relation-aware knowledge embedding layer \cite{Wang00LC19} is adopted to aggregate different types of entities in KG into items $V$ to enhance items' semantic information for constructing a knowledge-aware user-item graph $\textbf{G=(U, V, Y)}$.
Finally, we introduce a graph embedding layer with attention-aware mechanisms to embed users and items for main view construction.

\subsubsection{Multimodal Features Aggregation}
In this section, we  utilize pretrained neural networks to extract the multimodal entities of an item $v_j$ for constructing an associated knowledge graph $G_k=(h, r, t)$.
In particular, we use PNASNet \cite{LiuZNSHLFYHM18} to extract visual entities from images and preprocessed words \cite{LiuXLSH24} to extract text entities, respectively.
After all the entities of items have been extracted, we can construct an item–entity graph $G_k=(h, r, t)$.
The triplet $(h, r, t)$ defines the semantic relatedness between the head entity $h$ and the tail entity $t$ in relation type $r$, where the head entity $h$ and tail entity $t$ represent items in user-item graphs $G$ and a specific type of multimedia content of items, respectively.

To enhance items' semantic information, we incorporate KG as a comprehensive information network into items in user-item graphs.
Given the associated knowledge graph $G_k=(h, r, t)$ and a user-item graph $G=(U, V, Y)$, we employ a relation-aware knowledge embedding layer \cite{Wang00LC19} to aggregate different types of entities in KG into items $V$ to enhance items' semantic information for generating a knowledge-aware user-item graph $\textbf{G=(U, V, Y)}$, which can effectively capture of diverse relationships inherent in the connection structure of the KG. 
The relation-aware knowledge embedding layer between an item and its connected entities can be obtained by formula \ref{formula32}.

\begin{equation}
\label{formula32}
\mathbf{z_j}=Norm(z_j+\sum_{e\in N_j} a(e, r_{e,j}, j) \cdot z_e)
\end{equation}
{where} $z_j \in \mathrm{R^d} $ and $z_e \in \mathrm{R^d} $ represent the embeddings of an item $j$ and an entity $e$ to which it is connected, respectively.
$N_i$ represents the neighboring entities of an item $j$ by different types of relations $r_{e, j}$ in KG. We adopt function $Norm$ for normalization and $\mathbf{z_i}$  is enhanced embeddings of an item $j$.

To distinct semantics of relationships between item $i$ and entity $e$, we use $a(e, r_{e, j}, j)$ to estimate entity-specific and relation-specific attentive relevance, which can be obtained by formula \ref{formula33}.

\begin{equation}
\label{formula33}
a(e, r_{e, j},j)=\frac{exp(\sigma (r_{e, j}^\mathrm{T} \mathrm{W}[z_e||z_j]))}{\sum_{e \in N_j} exp(\sigma (r_{e, j}^\mathrm{T} \mathrm{W}[z_e||z_j]))}
\end{equation}
{where} $\mathrm{W \in R^{d \times 2d}}$ represents a parametric weight matrix, $\sigma$ is a nonlinear activation function \cite{Wang00LC19},  and $r_{e, j}^\mathrm{T}$ is an attention vector.
In the knowledge aggregation process, $a(e, r_{e, j},j)$ can distinct semantics of relationships between item $j$ and entity $e$ and the semantic information of an item $j$ can be enhanced by the KG.

\begin{figure*}
  \centering
  \includegraphics[width=0.74 \linewidth]{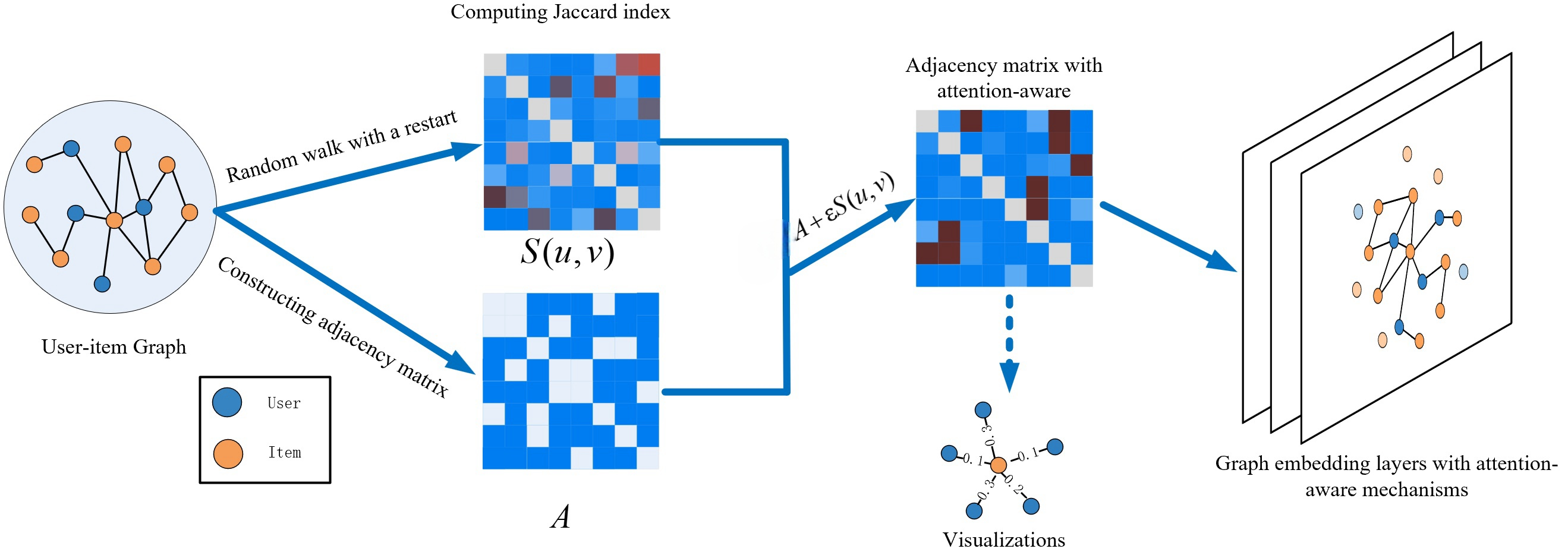}
    \vspace{-5pt}
  \caption{{Graph embedding layer with attention-aware mechanisms: the original user-item graph is used to construct an attention-aware matrix $S$,  which is constructed by computing the Jaccard index for the set of a user and an item sampled by a random walk with a restart strategy as the similarity of the user and item; The attention-aware matrix $S$ is then incorporated into adjacency matrix $A$ to construct a graph embedding layer with attention-aware mechanisms for learning the importance between users and items; a visual adjacency matrix with attention-aware is shown to clarify the graph embedding propagation layer on the node level with the contributions of each neighboring node.}}
  \label{attentionmatrix}
\end{figure*}

\subsubsection{Attention-aware User-item Graph Learning}
As discussed in Section \ref{introduction}, it is a challenge to design an attention-aware matrix by employing topology relationships of user-item graphs into GNN to learn the importance of user-item interactions for enhancing explanation and efficiency.
{Hence, we design a graph embedding layer with attention-aware mechanisms as shown in Figure \ref{attentionmatrix} to learn the importance of user-item interactions for generating representations of users $\textbf{U}$ and items $\textbf{V}$ after obtaining the user-item graph $\textbf{G=(U, V, Y)}$.}
Specifically, the graph embedding layers embed a user $\mathbf{u_i}$ and an item $\mathbf{v_j}$ as embedding vectors $\mathbf{z_i^u} \in R^d$ and $\mathbf{z_j^v} \in R^d$.
The embeddings of users $\textbf{U}$ and items $\textbf{V}$ can be defined as embedding matrices $\mathbf{Z^{(u)}} \in R^{I \times d}$ and $\mathbf{Z^{(v)}}\in R^{ J \times d}$, respectively.
Therefore, we remove feature transformation matrices and non-linear activation functions from GNN to construct a simplified graph embedding propagation layer with attention-aware mechanisms for generating node representations, which can be obtained by formula \ref{formula101}. 

\begin{equation}
\label{formula101}
\mathbf{z_i^{(u)}}=\mathbf{\bar{L}_{i,*}}\cdot \mathbf{Z^{(v)}}, \quad \mathbf{z_j^{(v)}}=\mathbf{\bar{L}_{*,j}}\cdot \mathbf{Z^{(u)}},
\end{equation}
{where} $\mathbf{z_j^{(v)}} \in R^d$ and $\mathbf{z_i^{(u)}} \in R^d$ define aggregated node features from neighbouring nodes to the central item $v_j$ and user $u_i$, respectively.
The $\mathbf{\bar{L}}$ represents a normalised Laplacian matrix with attention-aware and can be obtained by formula~\ref{formula320}.

\begin{equation}
\label{formula320}
\mathbf{\bar{L}}=\mathbf{D_{(u)}^{-\frac{1}{2}}}(\mathbf{A} + \xi \mathbf{S})\mathbf{D_{(v)}^{-\frac{1}{2}}}, 
\end{equation}
{where} $\mathbf{A}$ represents the adjacency matrix of user-item graphs. The diagonal degree matrice for users $\mathbf{U}$ and items $\mathbf{V}$ represent $\mathbf{D_{(u)}}$ and $\mathbf{D_{(v)}}$, respectively.

An attention-aware matrix $\mathbf{S}$ represents the similarity between user and item interactions and contributions can be controlled by the hyperparameter $\xi$, defined by formula~\ref{formula36}.

\begin{equation}
\label{formula36}
\mathbf{S}(u,v)=\frac{ |A_{M,R}(u) \cap  B_{M,R}(v)|}{|A_{M,R}(u) \cup  B_{M,R}(v)|}
\end{equation}
{where} $A_{M, R}(u)$ and $B_{M, R}(v)$ define the set that is sampled, which can be generated by a random walk with a restart strategy~\cite{ZhangSZZ23} from a starting node $u/v$ via the number of sampled paths $R$ and the length of the sampled paths $M$.
Therefore, the importance between a user $u$ and an item $v$ can be preserved by attention-aware matrix $\mathbf{S}(u,v)$.

In the graph embedding propagation process, attention-aware matrix $\mathbf{S}(u,v)$ can identify the importance between users and items.
Finally, we employ multiple embedding propagation layers to aggregate local neighbor information to refine user and item embeddings, which can be obtained by formula~\ref{formula340}.

\begin{equation}
\label{formula340}
\mathbf{z_{i,l+1}^{(u)}}=\sum_{v \in  N_u} \frac{\mathbf{z^v_{i,l}}}{\sqrt{|N_{u}| \cdot |N_{v}|}}, \quad \mathbf{z_{j,l+1}^{(v)}}=\sum_{u \in  N_v} \frac{\mathbf{z^u_{j,l}}}{\sqrt{|N_{u}| \cdot |N_{v}|}}
\end{equation}

In $l$-th embedding propagation layer, we define the embedding of a user $u_i$ and an item $v_j$ as $\mathbf{z_{i,l+1}^{(u)}}$ and $\mathbf{z_{j,l+1}^{(v)}}$, respectively.
We utilize the inner product between the final embedding of node $u_i$ and node $v_j$ to predict $u_i$’s preference towards $v_j$, as shown in Formula~\ref{formula35}.

\begin{equation}
\label{formula35}
y_{i,j}=\mathbf{z_{i}^{(u)\text{T}}}\mathbf{z_{j}^{(v)}}
\end{equation}

\begin{figure}
  \centering
  \includegraphics[width=0.80 \linewidth]{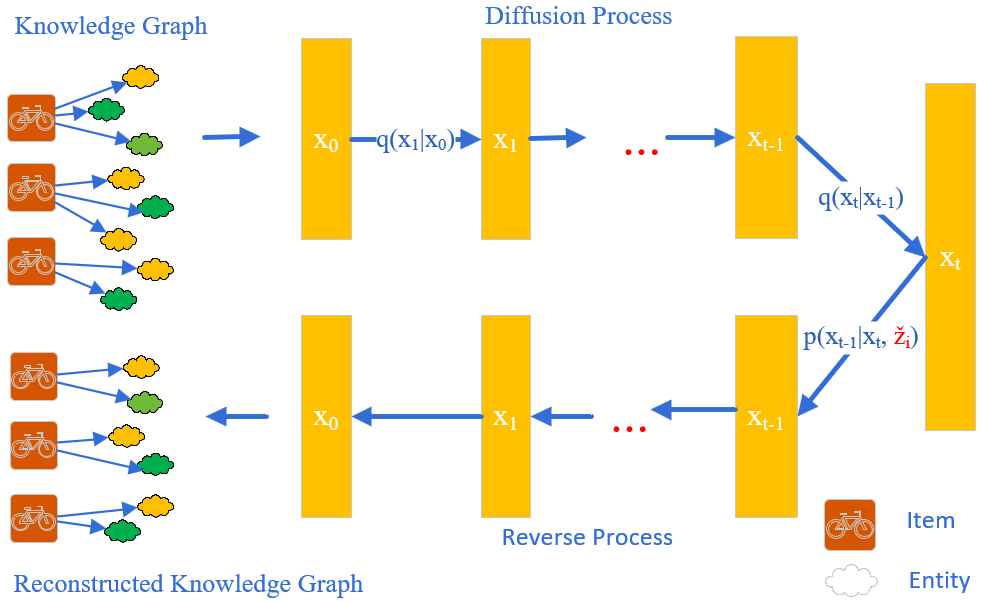}
    \vspace{-5pt}
  \caption{{Guided diffusion model: Diffusion process aims to add noise to corrupt the relationships between items and entities in the knowledge graph $G_k$; Reverse process aims to employ user embeddings with an edge connected to an item to guide the recovery of the original relationships between items and entities iteratively from pure Gaussian noise.}}
  \label{toyexampleDP}
\end{figure}

\subsection{Contrastive Learning with Guided Diffusion-enhanced}
As discussed in Section \ref{introduction}, some researchers have employed CL to generate a new knowledge graph for addressing data sparsity issues in user-item graphs.
Nevertheless, most existing approaches usually focus on simplistic random augmentation, which may raise noise and topic-irrelevant information between items and entities by multimodal noise.
Specifically, only a subset of the wealth of multimedia content is truly relevant in practical scenarios, resulting in generated contrastive views that may be contaminated.
Recently, some user-item graph contrastive learning approaches have introduced diffusion models to generate robust knowledge graphs, which can retain the relationships relevant to downstream tasks and alleviate noise issues.
However, most existing methods ignore using original knowledge graph relationships as supplementary content to guide the generation of strongly task-relevant node KG with less noise, which can't ensure that the generated KG can always bring benefits towards the task-relevant node knowledge.

{To solve the problems that have been identified above, we propose a guided diffusion model as shown in Figure \ref{toyexampleDP} to generate strongly task-relevant knowledge graphs with less noise for constructing a knowledge-aware contrastive view, which employs user embeddings with an edge connected to an item to guide the generation of strongly task-relevant knowledge graphs for enhancing the item's semantic information.}
Furthermore, the generated KG structure is adjusted by top-$q$ relations between items and entities that are strongly task-relevant, which can preserve the informative structure of the reconstructed knowledge graph with less noise.
In more detail, we add noise to corrupt the relationships between items and entities in the knowledge graph $G_k$ in the diffusion phase.
In the reverse process, we recover the original relationships to identify true relationships between items and entities iteratively from pure Gaussian noise.
From restored relation probabilities, we can reconstruct strongly task-related subgraph $\hat{G_k}$ and adjust the structure of $\hat{G_k}$ with less noise by top-$q$ relations for constructing contrastive views.

\subsubsection{Diffusion Process}
We perform reverse to identify task-relevant relationships with less noise between items and entities for reconstructing strongly task-related subgraph $\hat{G_k}$.
Therefore, we add noise to corrupt the relationships $x_0$ between items and entities in the original knowledge graph $G_k$ in the diffusion phase, which can be obtained by formula \ref{formula330}.

\begin{equation}
\label{formula330}
q(x_t|x_0)=N(x_t; \sqrt{\bar{a_t}}x_0,(1-\bar{a_t})\mathbf{I}),\quad \bar{a_{t}}= \prod_{t^{'}=1}^t(1-\beta_{t^{'}})
\end{equation}
where $\mathbf{I}$ represents an identity matrix, $t\in \{1 \dots T\}$ represents the diffusion step, and $N$ represents the Gaussian distribution. 
In each step $t$, we adopt $\beta_t \in (0,1)$ to control the scale of the Gaussian noise added.
Original relationships between an item $j$ and an entity $e$ to which it is connected can be defined as initial state $x_0=\mathbf{r_j}$, which can be got by a matrix, as displayed in formula~\ref{formula322}:

\begin{equation}
\label{formula322}
\mathbf{r_j}=[ r_j^0,r_j^1,\dots,r_j^{|\varepsilon |-1} ]
\end{equation}
where $r_j$ represents an item $j$ that has relations with entities in the entity set $\varepsilon$.
If $r_j^e=1$, item $j$ has a relation with entity $e$, and vice versa.
As $T \to \infty$, the state $x_T$ converges towards a standard Gaussian distribution. 

\subsubsection{Reverse Process}
In the reverse process, we focus on recovering relationships $x_0$ iteratively from a pure Gaussian noise $x_T$.
Specifically, we employ user embeddings with an edge connected to an item as supplementary content to guide the generation of strongly task-relevant node KG. 
The diffusion model adopts neural networks to remove the added noises by learning to recover $x_{t-1}$ from $x_t$, which can be obtained by formula \ref{formula382}.

\begin{equation}
\label{formula382}
p_{\theta}(x_{t-1}|x_t, \tilde{z_i})= N(x_{t-1}; u_{\theta}(x_t,t,{z_i}),\Sigma_{\theta}(x_t,t,\tilde{z_i}))
\end{equation}
where $\Sigma_{\theta}(x_t,t,\tilde{z_i})$  and $u_{\theta}(x_t,t,\tilde{z_i})$ represent covariance of Gaussian distribution and mean, which can be got by utilizing neural networks parameterized with $\theta$.
$\tilde{z_i}$ define user embeddings with an edge connected to an item of the original knowledge graph, which is utilized to guide the generation of strongly task-relevant node KG and can be obtained by  formula \ref{formula38012}.

\begin{equation}
\label{formula38012}
\tilde{z_i}=Avg(\sum_{i \in N_j} z_i)
\end{equation}
where $z_i$ define user embeddings of the original knowledge graph, $N_j$ represents the neighboring users with an edge connected to an item $j$, and $Avg$ defines the average operation.
We can reparameterize the mean $u_{\theta}$ to learn the added noise in time step $t$ by neural networks,  as displayed in formula \ref{formula383}.

\begin{equation}
\label{formula383}
u_{\theta}(x_t,t,\tilde{z_i})=\frac{1}{\sqrt{\bar{a_t}}}(x_t-\frac{\beta_{t}}{\sqrt{1-\bar{a_t}}}{\epsilon_{\theta}}(x_t,t,\tilde{z_i}))   
\end{equation}

We adopt a Multi-Layer Perceptron to implement $\epsilon_{\theta}(x_t,t,\tilde{z_i})$. Specifically, we adopt the $x_t$, step embedding $t$, and user embeddings $\tilde{z_i}$ as inputs to predict $\hat{x}_0$.
Furthermore, we utilize user embeddings $\tilde{z_i}$ to guide $\hat{x}_0$ embedding reconstruction.
Thus, the user embeddings adjust the embedding of the $\hat{x}_0$, which can guarantee that the generated data can always bring benefits towards the diffusion augmentation model.
Ultimately, maximum the ELBO \cite{JiangYXH24} of the likelihood of $x_0$ is adopted to update model parameters.

\subsubsection{Contrastive View Generation}
\label{topq}
We focus on generating a contrastive view with minor noise for containing strongly task-relevant node knowledge in this section.
Therefore, we use $\hat{x}_0$ to adjust the KG structure for reconstructing strongly task-relevant knowledge graph $\hat{G_k}$ with less noise after getting the reconstructed $\hat{x}_0$.
In more detail, we select top-$q$ relations between items $j$ and entities $e$ from $\mathbf{\hat{r}_j}$ that are strongly task-relevant to modify KG structure, which can preserve the informative structure of the reconstructed knowledge graph $\hat{G_k}$ with less noise.
Then, we aggregate entities in reconstructed knowledge graph $\hat{G_k}$ into items $V$ by the formula \ref{formula32}, which can enhance items' semantic information for generating a contrastive view $\mathbf{\hat{G}=(\hat{U}, \hat{V}, \hat{Y})}$.
Finally, embeddings $\mathbf{\hat{z}_i^{(u)}} \in R^d$ and $\mathbf{\hat{z}_j^{(v)}}\in R^d$ for user $\hat{u}_i$ and item $\hat{v}_j$ are generated by the formula \ref{formula340} for data augmentation, respectively.

\subsection{Model Training}
In this section, we utilize InfoNCE loss~\cite{abs-1807-03748} to contrast augmented view embeddings with main-view embeddings for model parameters training and the contrastive loss for users $U$ defines as formula~\ref{contrastive-1}.

\begin{equation}
\label{contrastive-1}
L^{(u)}_{s}=\sum_{u \in U}-\log\frac{\exp{((s(\mathbf{z^{(u)}},\mathbf{\hat{z}^{(u)}})/\tau)}}{\sum_{v\in U}\exp{((s(\mathbf{z^{(u)}},\mathbf{\hat{z}^{(v)}})/\tau)}},
\end{equation}

Where $s(\cdot)$ and $\tau$ represent the cosine similarity and the temperature, respectively.
The $(\mathbf{z^{(u)}},\mathbf{\hat{z}^{(u)})}$ represent the same nodes in different views as positive pairs, while $(\mathbf{z^{(u)}},\mathbf{\hat{z}^{(v)}})$ ($v\in U$) represents any two different nodes in different views as negative pairs.
We represent item $V$ contrastive loss $L^{(v)}_{s}$ in the same way.
We jointly optimize the main objective function with the contrastive loss for model parameters training, which can be obtained by formula~\ref{formula42}:

\begin{equation}
\label{formula42}
L=L_r+\theta_1(L_s^{(u)}+L_s^{(v)})+\theta_2 \cdot ||\Theta||_2^2,
\end{equation}

We employ $\theta_2$ to control the contribution of the model parameters $\Theta$ and $\theta_1$ to control the contribution of contrastive loss.
We define the main objective function as $L_r$ and can be described as formula \ref{formula403}.
The $y_{u, i}$ represents the predicted scores for a pair of positive item $v$ of user $u$, while $y_{u, j}$ represents the predicted scores for a pair of negative item $v$ of user $u$.

\begin{equation}
\label{formula403}
L_r=\sum_{(u,i,j) \in O}  -log(y_{u, i}-y_{u, j})
\end{equation}

\smallskip\noindent
\textbf{Interpretability Analysis:} 
We compute the Jaccard index \cite{WangNB23} for the set of a user and an item sampled by a random walk with a restart strategy as the similarity of the user and item to build the attention-aware matrix $\mathbf{S}$, which can preserve the importance between users and items and generate attention-aware node feature aggregation.
Intuitively, the more similar the user $u$ and the item $v$, the larger the $\mathbf{S}(u, v)$ value.
The mechanism can enhance the interpretability of the KDiffE model compared with adaptive training attention weight mechanisms.
In more detail, the graph embedding propagation layer aggregates information from the node and its neighboring nodes, and attention-aware matrix $\mathbf{S}$ clarify the graph embedding propagation layer on the node level with the contributions of each neighboring node.
Therefore, the interpretability of the KDiffE model can be enhanced by the attention-aware $\mathbf{S}$.


\section{Evaluation}
In this section, extensive experiments are performed on three public datasets to estimate our proposed KDiffE model.

\subsection{Experimental Settings}
\label{experimental}
\subsubsection{Datasets}
\label{dataset}
We chose the TikTok, Amazon-Baby, and Amazon-Sports datasets to estimate our proposed KDiffE model.
Table \ref{tab1} shows the detailed statistical information of the three datasets, and we represent visual, acoustic, and textual features as V, A, and T, respectively.

\begin{table}
\caption{{Datasets statistical properties}}
\begin{center}
\begin{tabular}{|c|ccc|c|}
\hline
Datasets & User & Item & Interactions & Modality \\

\hline
TikTok &  9,319& 6,710 & 59,541 & V \ A \ T \\

 Amazon-Baby &  19,445 &7,050&139,110  & V \ T\\

 Amazon-Sports &  35,598 & 2,18,357 &256,308   & V \ T\\

\hline

\end{tabular}
\label{tab1}
\end{center}
\end{table}

\begin{table*}
  \caption{{Recommendation Performance on three datasets in terms of Recall@20 and NDCG@20}}
  \centering
  \label{SRPerformance}
  \begin{tabular}{ccccccl}
    \toprule
Datasets & \multicolumn{2}{c}{TikTok} & \multicolumn{2}{c}{Amazon-Baby} & \multicolumn{2}{c}{Amazon-Sports} \\
    \cline{1-7}
    Baselines & Recall@20& NDCG@20& Recall@20& NDCG@20& Recall@20&NDCG@20\\
    \cline{2-7}
     SGL &0.060 &0.024 & 0.068& 0.030& 0.078& 0.036\\
      NCL & 0.066 &0.027 & 0.070& 0.031& 0.077& 0.035\\
        HCCF & 0.066 &0.027 & 0.071& 0.031&0.078& 0.036\\
        CLCRec & 0.062 &0.026 & 0.061& 0.028& 0.065& 0.030\\
           MMGCL & 0.080 &0.033 & 0.076& 0.033& 0.088& 0.041\\
            SLMRec & 0.085 &0.035 & 0.077& 0.033& 0.083& 0.038\\
             BM3 & 0.096 &0.040 & 0.084& 0.036& 0.098& 0.044\\
             MGCL &  0.109& 0.040& 0.087& 0.038& 0.100& 0.044\\
             MHGCF &  0.100&0.043 & 0.091& 0.039& 0.097&0.040 \\
             DiffKG & 0.099 & 0.044& 0.087&0.037 & 0.095& 0.042\\
              \cline{1-7}
               \textbf{KDiffE} & \textbf{0.112}& \textbf{0.046}&\textbf{0.095} &\textbf{0.040} &\textbf{0.102} & \textbf{0.046}\\

    \bottomrule
  \end{tabular}
\end{table*}

\begin{itemize}
  \item \textbf{TikTok:} It stores a large amount of short-form video content, which captures user interactions with rich visual, acoustic, and textual features. 
  \item \textbf{Amazon-Baby:} It is a multimedia dataset collected from the Amazon platform with rich visual and textual features, which includes 19,445 users, 7,050 items, 139,110 interactions. 
    \item \textbf{Amazon-Sports:} It is a multimedia dataset collected from the Amazon platform with rich visual and textual features, which includes 35,598 users, 2,18,357 items, 256,308 interactions. 

\end{itemize}

\subsubsection{Baselines}
\label{baselines}
In this section, we select two types of baselines, including CL-based recommendation models (SGL~\cite{WuWF0CLX21}, NCL~\cite{LinTHZ22}, and HCCF~\cite{XiaHXZYH22}) and multi-modal recommendation models (CLCRec~\cite{WeiWLNLLC21}, MMGCL~\cite{Yi0OM22}, SLMRec~\cite{TaoLXWYHC23}, BM3~\cite{ZhouZLZMWYJ23}, MGCL~\cite{LiuXGSQH23}, MHGCF~\cite{LiuXLSH24}, and DiffKG~\cite{JiangYXH24}), to estimate the effectiveness of the KDiffE model.

\begin{itemize}

  \item \textbf{SGL~\cite{WuWF0CLX21}: } It adopts random data augmentation operators to enhance contrastive learning signals for recommendations.
    \item \textbf{NCL~\cite{LinTHZ22}: } It generates positive contrastive pairs by identifying neighboring nodes identifies neighboring nodes to construct contrastive views for recommendations.
   \item \textbf{HCCF~\cite{XiaHXZYH22}:} It enhances hypergraph neural networks by cross-view contrastive learning paradigms to learn local and global collaborative relations for recommendations.
   \item \textbf{CLCRec~\cite{WeiWLNLLC21}:} It adopts multi-modal features using contrastive learning to enrich item embeddings for handling the item cold-start problem in MMRec.   
   \item \textbf{MMGCL~\cite{Yi0OM22}:} It enhances multi-modal representations by modality edge dropout and modality masking in a self-supervised learning manner.
   \item \textbf{SLMRec~\cite{TaoLXWYHC23}:} It adopts multi-modal pattern uncovering and noise perturbation over features to enhance data for multi-modal content.
   \item \textbf{BM3~\cite{ZhouZLZMWYJ23}:} It adopts self-supervised learning to eliminate the need for randomly sampled negative samples in MMRec.
   \item \textbf{MGCL~\cite{LiuXGSQH23}:} It learns visual preference clues and textual preference clues using a CL-based strategy in MMRec.
  \item \textbf{MHGCF~\cite{LiuXLSH24}: } It constructs a knowledge graph to enhance items' semantic information by extracting collaborative signals, content-level preferences, and semantic-level preferences.
    \item \textbf{DiffKG~\cite{JiangYXH24}: } It adopts diffusion models and graph contrastive learning to learn a knowledge graph for enhancing items' semantic information for recommendations.

\end{itemize}

\subsubsection{Parameter settings}
\label{settings}
We set $\theta_1 = 1e^{-2}$ and $\theta_2 = 1e^{-5}$, which control the contribution of the contrastive losses and the model parameters, respectively. The number of sampled paths $R$ is set to 12, and the length of the sampled paths $M$ is set to 50. we set the hyperparameter to control the contribution of the attention-aware matrix $\xi=0.7$, the number of steps $t=10$, the temperature parameter $\tau$ search from $\{0.5, 0.7\}$, and the parameter top-$q$ relations for adjusting the strongly task-relevant KG structure $q=1$.
We employ the Recall@N and NDCG@N~\citep{LiuXLSH24} with $N=20$ to evaluate our KDiffE model.
We fine-tune the baseline to the optimal value to ensure fair comparisons and perform 10 times experiments showing average metrics. 
The experiments are conducted on the Ubuntu 22.04.4 operating system with a Intel(R) Xeon(R) Silver 4310 CPU @ 2.10GHz machine, 1024 memory, NVIDIA Corporation Device 2684, and Python 3.11.

\subsection{Recommendation Performance}
\label{Performance}
The experimental results demonstrate that our KDiffE model performs the best performance, as shown in Table \ref{SRPerformance}.
The SGL, NCL and HCCF primarily employ CL-based paradigms to embed nodes, which ignore utilizing rich multimedia content to enhance historical user-item interaction information, resulting in lower performance than MMRec-based models.
The CL-based multi-modal recommendation models employ rich multimedia content to enhance node semantic information, such as BM3 and MGCL, which improves recommendation performance.
DiffKG introduces diffusion models and graph contrastive learning to learn multimedia content for enhancing items' semantic information, resulting in significantly improved performance and beats most models. 
Nevertheless, still beaten by our KDiffE model.
Our KDiffE model adopts contrastive learning architecture and designs an attention-aware matrix to identify the importance between users and items, which can generate attention-aware node feature aggregation and alleviate the sparsity problem.
Furthermore, we propose a guided diffusion model to generate a knowledge-aware contrastive view, which can generate a task-relevant node KG with less noise for enhancing node semantic information.
Thus, our model performs the best performance.

\begin{table}
  \caption{{Ablation Study on key components of KDiffE}}
  \label{tab11a}
  \begin{tabular}{c|ll|ll|ll}
    \hline
    Datasets &\multicolumn{2}{c|}{TikTok} & \multicolumn{2}{c|}{Amazon-Baby} & \multicolumn{2}{c}{Amazon-Sports}\\
        \hline

     Variants&Recall& NDCG&Recall&NDCG&Recall&NDCG\\
    \hline
    KDiffE\_1 &0.107 & 0.041&0.094 &0.040&0.101&0.045\\
    KDiffE\_2 & 0.110&0.045 & 0.093&0.040&0.101&0.045\\
    KDiffE\_3 &0.107 &\textbf{0.046} &0.091 &0.038&0.098&0.043\\
    \hline
    \textbf{Ours} & \textbf{0.112}& \textbf{0.046}&\textbf{0.095} &\textbf{0.041}&\textbf{0.102}&\textbf{0.046}\\
    \hline

  \end{tabular}
\end{table}

\subsection{Ablation Study}
\label{ablation}
In this section, we organize an ablation study to demonstrate the effectiveness of the attention-aware matrix, guided diffusion mechanism, and contrastive learning modules. We execute 5 experiments to display the average Recall@20 and NDCG@20 values.

\subsubsection{Effectiveness of Attention-aware Matrix}
\label{Attention-Matrix}
To demonstrate the contribution of the attention-aware matrix $S$ module, we remove the $S$ from GNN denoted as KDiffE\_1 and report the average Recall@20 and NDCG@20 values.
As displayed in Table~\ref{tab11a}, the experimental results demonstrate the effectiveness of the attention-aware matrix.
Especially on the TikTok dataset, the average NDCG of KDiffE is 0.5\% higher than KDiffE\_1 and the average Recall value is 0.5\% higher than KDiffE\_1.
One possible explanation is that KDiffE\_1 aggregates node interactions on average, failing to identify the importance between users and items.
Instead, the KDiffE model builds an attention-aware matrix to learn the importance between users and items, which can generate attention-aware node feature aggregation and improve recommendation performance.



\subsubsection{Effectiveness of Guided Diffusion Mechanism}
\label{Guided-Matrix}
We utilize user embeddings $\tilde{z_i}$ to guide the generation of task-relevant node KG, as displayed in the formula \ref{formula382}.
To verify its contribution, we remove user embeddings $\tilde{z_i}$  from formula \ref{formula382} denoted as KDiffE\_2 and report the average Recall@20 and NDCG@20 values.
As displayed in Table~\ref{tab11a}, the experimental results demonstrate the effectiveness of the guided diffusion mechanism, which can guide the generation of task-relevant knowledge graphs for enhancing the item's semantic information and improving model performance.


\subsubsection{Effectiveness of Contrastive Learning}
\label{Guided-CL}
To verify the contribution of contrastive learning, we remove the $L_s^{(u)}$ and $L_s^{(v)}$ losses from ${L}$ denoted as KDiffE\_3 and report the average Recall@20 and NDCG@20 values.
As illustrated in Table \ref{tab11a}, although the average NDCG value of KDiffE is equivalent to that of KDiffE\_3 on the TikTok dataset, the experimental results indicate that the Recall value demonstrates significant effectiveness across all three datasets, thereby enhancing data quality and subsequently improving performance."


\begin{figure}
  \centering
  \includegraphics[width=1.0 \linewidth]{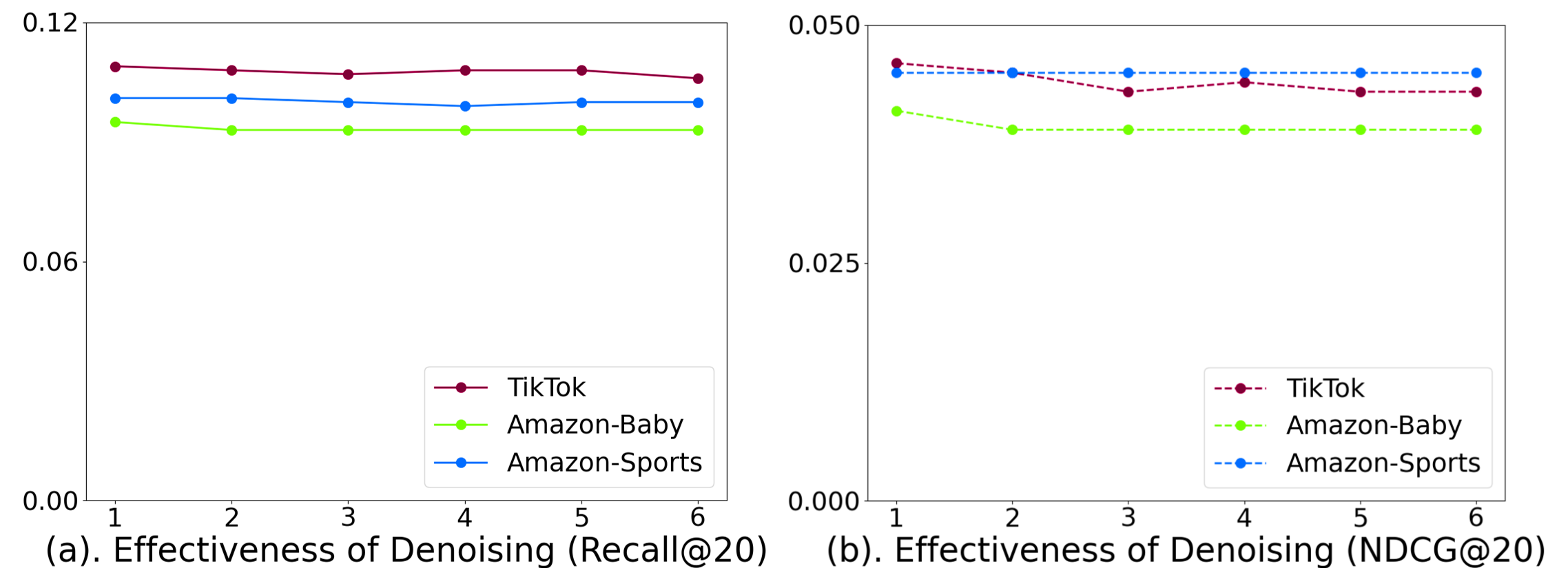}
    \vspace{-5pt}
  \caption{Effectiveness of Denoising}
  \label{EffDenoising}
\end{figure}

\subsection{Effectiveness of Denoising}
\label{Denoising}
As discussed in Section \ref{topq}, we select top-$q$ relations between items $j$ and entities $e$ from $\mathbf{\hat{r}_j}$ that are strongly task-relevant to modify KG structure, which can generate a contrastive view with less noise for containing task-relevant node knowledge.
In particular, the smaller the $q$ value, the less the increased task-relevant relations, and the less noise introduced, and vice versa.
Hence, we can adjust the $q$ value to control the introduction of noise.
In the experiment, we found that with the parameter $q$ increases, the performance decreases, as shown in Figure~\ref{EffDenoising}.
Our model can perform the most satisfactory performance when $q=1$. 
As the $q$ continues to increase, the performance decreases. 
The reason may be that additional noise information is introduced, which worsens the recommendation performance.
Therefore, satisfactory recommendation performance can be obtained by adjusting the $q$ value to control the introduction of noise.

\begin{figure}
  \centering
  \includegraphics[width=1.0 \linewidth]{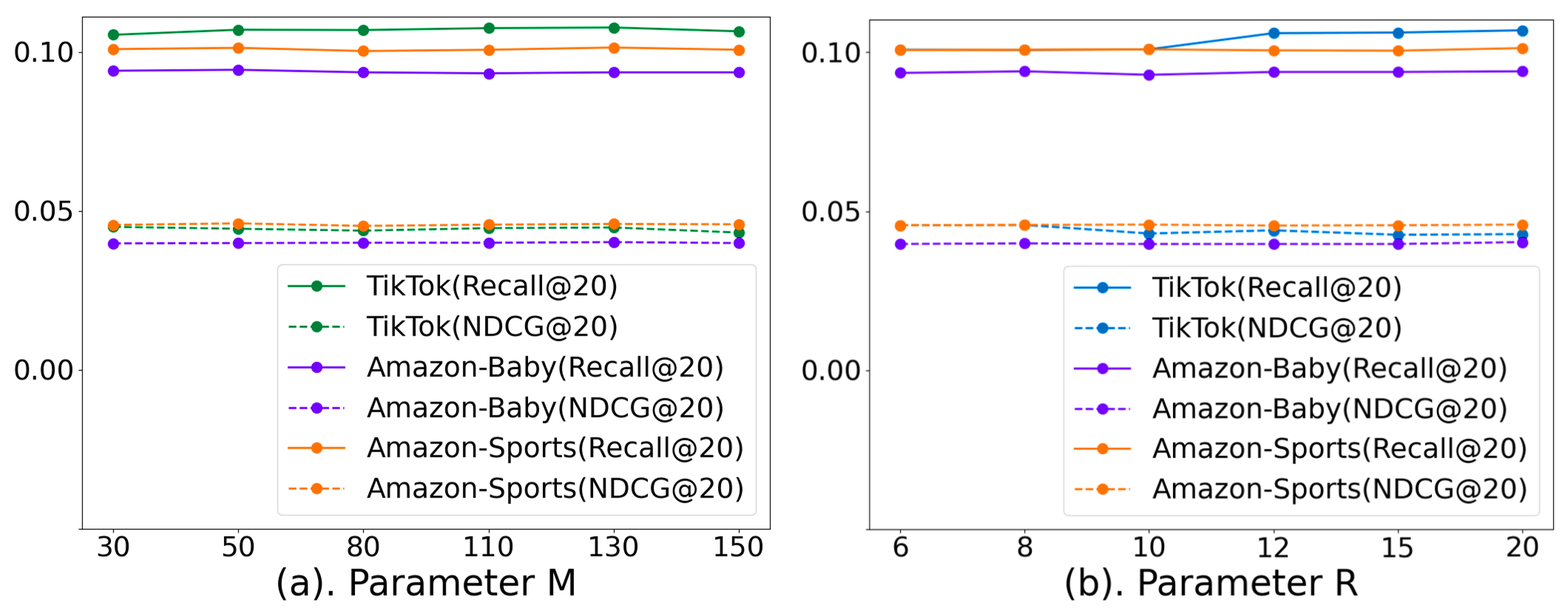}
    \vspace{-5pt}
  \caption{Hyperparameter analysis for $M$ and $R$}
  \label{EffectivenessMNP}
\end{figure}

\subsection{Hyper-parameter Analysis}
\label{Hyper}
In this section, we conduct hyperparameter analysis. 
Particularly, we estimate how different the number of sampled paths $R$, the length of the sampled paths $M$, the hyperparameter to control the contribution of the attention-aware matrix $\delta$, the number of steps $t$, the temperature parameter $\tau$, the parameter to control the contribution of the contrastive loss $\theta_1$, {and the parameter top-$q$ relations} can impact the recommendation performance.

\subsubsection{Parameter $M$ and $R$ }
We analyze both parameters together because both parameters jointly decide the sampling size.
We set the parameter $M$ to $\{30, 50, 80, 110, 130, 150\}$ and the parameter $R$ to  $\{6, 8, 10, 12, 15, 20\}$ to verify the recommendation performance of our KDiffE model.
The experimental results displayed in Figure~\ref{EffectivenessMNP} show that satisfactory  performance is achieved when $M=50$ and $R=12$.
As the $M$ and $R$ continue to increase, the performance remains unchanged or increases slightly. 
From the experimental results, we found that the model performance was insensitive to both parameters and a smaller value can achieve satisfactory performance. Considering the computational efficiency, we set $M=50$ and $R=12$.

\begin{figure}
  \centering
  \includegraphics[width=1.0 \linewidth]{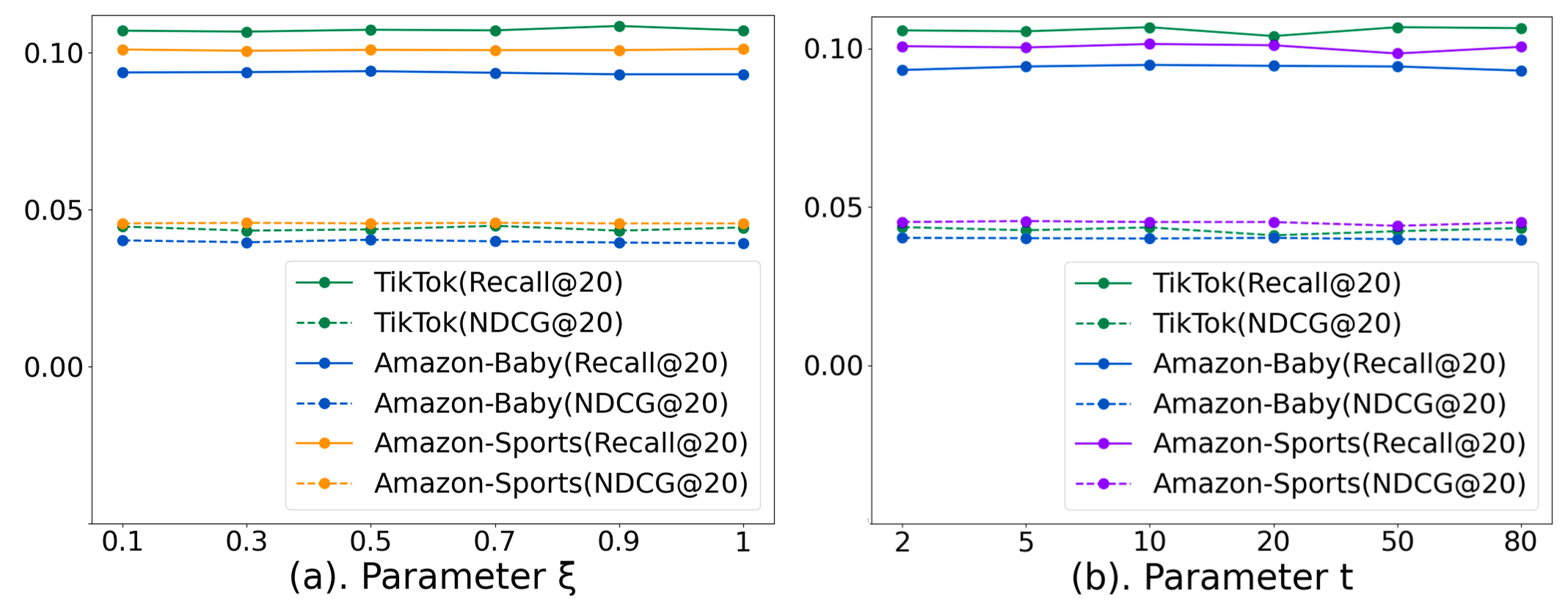}
    \vspace{-5pt}
  \caption{Hyperparameter analysis for $\xi$ and $t$}
  \label{Effectivenessktd}
\end{figure}

\subsubsection{Parameter $\xi$}
\label{historical}
To verify the parameter $\xi$, we search from $\{0.1, 0.3, 0.5, 0.7, 0.9, 1\}$ to evaluate the recommendation performance of our KDiffE model, which controls the contribution of the attention-aware matrix.
The experimental results displayed in Figure~\ref{Effectivenessktd}a that our model achieves the best performance varies by dataset and the satisfactory performance is achieved in three datasets when $\xi > 0.5$. A potential explanation is that the contribution of the attention matrix should be calibrated according to the sparsity characteristics of the dataset.
As the $\xi$ continues to increase, the performance remains unchanged or increases slightly. 
Considering the generalization ability of our model, we set $\xi=0.7$.

\begin{figure}
  \centering
  \includegraphics[width=1.0 \linewidth]{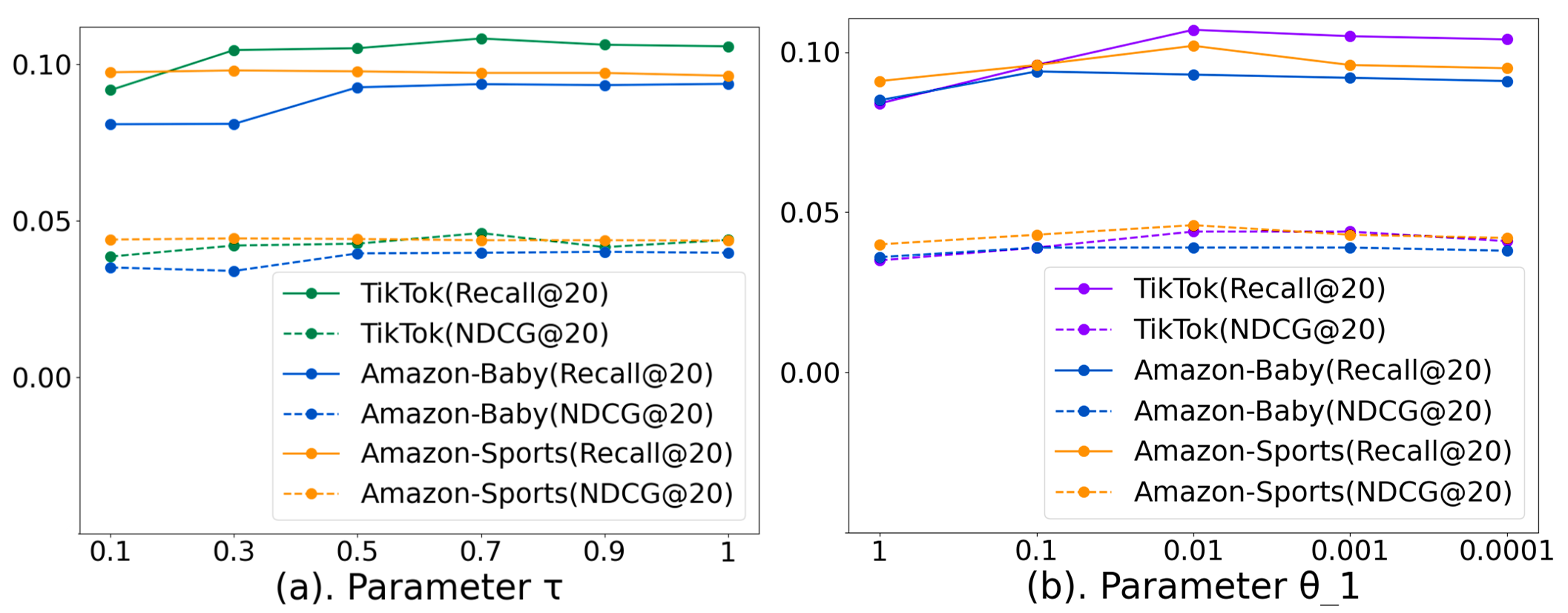}
    \vspace{-5pt}
  \caption{Hyperparameter analysis for $\tau$ and $\theta_1$}
  \label{Effectivenesskto}
\end{figure}

\subsubsection{Parameter $t$}
\label{historicalkt}
We set the parameter $t$ to 2, 5, 10, 20, 50, and 80 to verify the recommendation performance of our KDiffE model.
As the $t$ increases, the performance increases. 
When $t=10$, our model can obtain satisfactory results as shown in Figure~\ref{Effectivenessktd}b.
As the $t$ continues to increase, the performance decreases or remains unchanged.
Nevertheless, as the $t$ increases, the computational cost of our KDiffE model will increase.
Considering the balance between computational cost and performance, we set the $t=10$.

\subsubsection{Parameter $\tau$}
\label{historicalktt}
To verify the parameter $\tau$, we search from $\{0.1, 0.3, 0.5, 0.7, 0.9, 0.1\}$ to evaluate the recommendation performance of our KDiffE model.
The experimental results are shown in Figure~\ref{Effectivenesskto}a, the model achieves the best performance varies by datasets.
When $\tau$ in the range $[0.5, 0.7]$, our model can obtain satisfactory results as shown in Figure~\ref{Effectivenesskto}a.
As the $\tau$ continues to increase, the performance decreases or remains unchanged.

\subsubsection{Parameter $\theta_1$}
\label{historicalktto}
To verify the parameter $\theta_1$, we search from $\{1, 1e^{-1}, 1e^{-2} ,1e^{-3},1e^{-4}\}$ to evaluate the recommendation performance of our KDiffE model, which controls the contribution of the contrastive loss.
The experimental results are shown in Figure~\ref{Effectivenesskto}b, the best performance is achieved in three datasets when $\theta_1=1e^{-2}$.
As the $\theta_1$ continues to increase, the performance decreases. 
In the experiment, we found that the larger the $\theta_1$ values, the larger the contribution to contrastive learning, but the worse the performance.
The reason may be that a larger value makes the model pay too much attention to the contrastive learning task and reduces the focus on the main task, resulting in decreased performance.

\begin{figure}
  \centering
  \includegraphics[width=0.9 \linewidth]{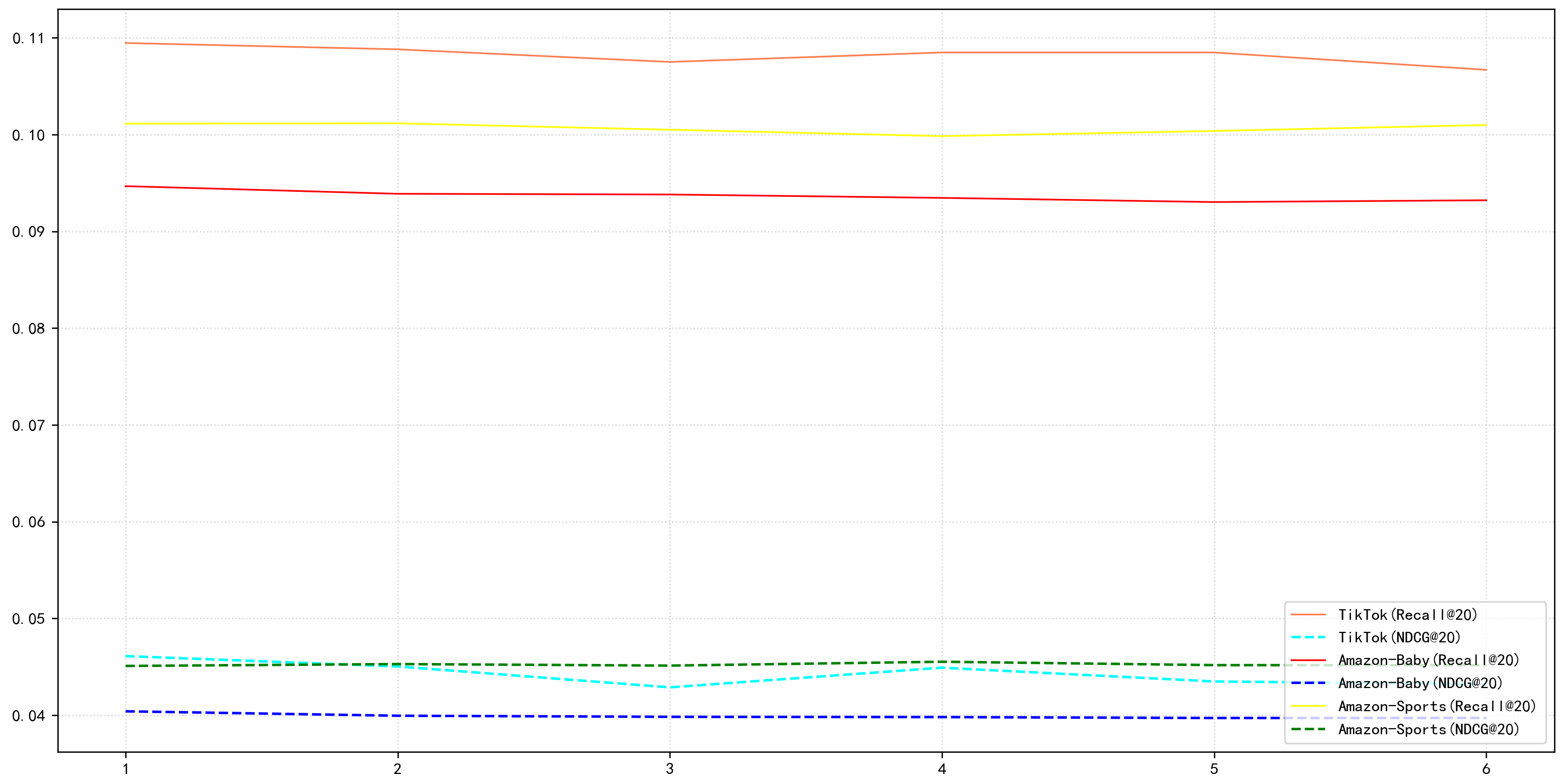}
    \vspace{-5pt}
  \caption{Hyperparameter analysis for $q$ }
  \label{Hyperparameterq}
\end{figure}

\subsubsection{Parameter $q$}
\label{historicalq}
{We set the parameter $q$ to 1, 2, 3, 4, 5, and 6 to verify the recommendation performance of our KDiffE model.
When $q=1/2$, our model can obtain satisfactory results as shown in Figure~\ref{Hyperparameterq}.
As the $q$ continues to increase, the performance decreases or remains unchanged.
Nevertheless, as the $q$ increases, the computational cost of our KDiffE model will increase.
Considering the balance between computational cost and performance, we set the $q=1$.}

\subsection{Scalability Analysis}
\label{efficiency}

\begin{figure}
  \centering
  \includegraphics[width=1 \linewidth]{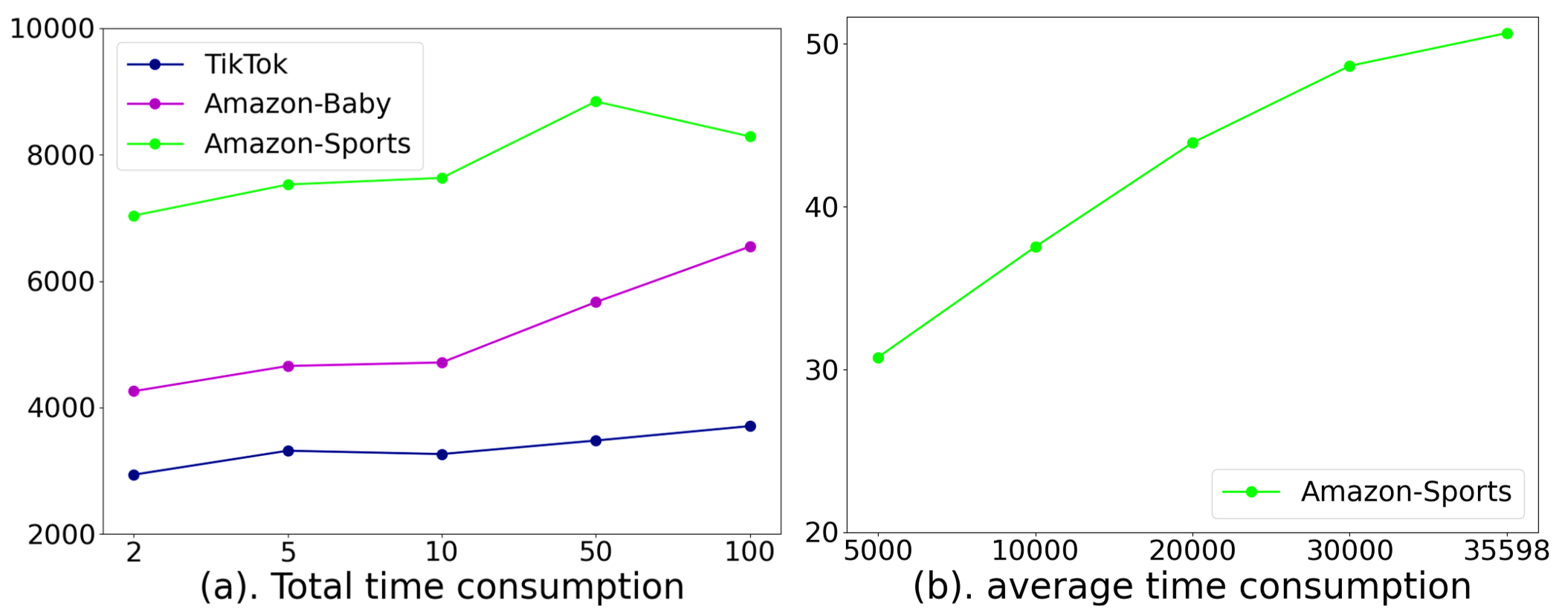}
    \vspace{-5pt}
  \caption{Scalability Analysis}
  \label{Scalabilitys}
\end{figure}

Diffusion-based models usually suffer from undesirable time consumption by different diffusion step lengths.
We show the time consumption of the KDiffE model at different diffusion steps on three multi-modal recommendation datasets to evaluate the efficiency.
As shown in Figure \ref{Scalabilitys}a, our model can perform satisfactorily at a small $t$-value, i.e. $t=10$. 
{In addition, since MHGCF~\cite{LiuXLSH24} constructs a knowledge graph to enhance items' semantic information in MMRec, while MGCL~\cite{LiuXGSQH23} learns visual preference clues and textual preference clues using a CL-based strategy in MMRec, we select both models, which do not employ diffusion, for comparison with our model in the time complexity analysis.
We define $M$ for the route length of the sampled paths, $R$ for the number of sampled paths, $L$ for the number of GNN, $d$ for the dimensionality, $s$ for the number of training epochs, $I$ and $J$ for the number of nodes $U$ and nodes $V$, $|Y|$ for the number of interactions in user-item graphs, $T$ for the number of diffusion step, $B$ for the number of nodes contained in a single batch, respectively. 
The training complexity of MGCL is close to $O((4|Y|\times L +6+4(B+1))\frac{ds|Y|}{B})$ and MHGCF is close to 
$O(2(|Y|\times (L+2))ds(\frac{|Y|}{B}))$. For our KDiffE model, the GCN with $L$ layer takes $O(L\times |Y| \times d)$, the attention-aware matrix $S$ takes $O ( R\times M\times(I+J))$, the contrastive learning paradigm takes $O(B \times L \times (I+J) \times d)$, and the guided diffusion model for generating task-relevant node knowledge graph $\hat{G_k}$ takes $O(|Y| \times d^2 \times T)$}.
Thus, we can infer that the training process of our KDiffE model will not be the bottleneck of model optimization.

Furthermore, graph contrastive learning models usually suffer a high computational cost due to constructing extra views.
The Amazon-Sports dataset contains more users than others, thus we utilize it for scalability analysis.
We set the number of users to 5,000, 10,000, 20,000, 30,000 and all nodes, respectively, to estimate the scalability of our model on the Amazon-Sports dataset.
As shown in Figure \ref{Scalabilitys}b, our KDiffE model takes about 28.34, 43.38, 46.00, 48.19 and 53.54 seconds per epoch on average with different settings for the number of users.
Thus, we can infer that as the number of users increases, the MHDiff model increases the computational cost linearly, and it is suitable for large-scale networks.

\section{Conclusion}
In this paper, we have introduced an effective augmentation method KDiffE to the graph contrastive learning framework for multimedia recommendation.
Especially, the attention-aware matrix is built by a random walk with a restart strategy to learn the importance between users and items for generating attention-aware node feature aggregation, which can improve computational efficiency and interpretability compared with based on adaptive training attention-weight models.
Then, we propose a guided diffusion model to generate a knowledge-aware contrastive view, which can generate a strongly task-relevant node KG with less noise for enhancing node semantic information.
Extensive experiments on three multimedia datasets reveal the effectiveness of our KDiffE and its components on different state-of-the-art baselines.
Our future work will explore the semantic relationships to guide knowledge graphs embedding reconstruction for improving multimedia recommendation performance.



\bibliographystyle{plain}

\vfill

\end{document}